\documentclass{jfmg}

\usepackage{graphicx}
\usepackage{amssymb}
\usepackage{amsfonts}
\usepackage{natbib}
\usepackage{pstool}
\usepackage[dvipsnames]{xcolor}
\usepackage{verbatim}
\usepackage{pifont}
\usepackage{psfrag}
\usepackage{amsmath}
\usepackage{appendix}
\usepackage{mathtools}
\usepackage[normalem]{ulem}

\graphicspath{ {./figures/} }

\newcommand{\ajs}[1]{{{\color{red} #1}}}
\newcommand{\blue}[1]{{{\color{blue} #1}}}

\title[Wall-wake laws for the mean velocity and the turbulence]{Wall-wake laws for the mean velocity and the turbulence}

\author[A. J. Smits]%
{Alexander J. Smits}

\affiliation{Department of Mechanical and Aerospace Engineering, 
Princeton University, \break
Princeton, NJ, USA\\[\affilskip]
}

\pubyear{2023}
\volume{}
\pagerange{}

\begin{document}

\maketitle

\begin{abstract}
A new wall-wake law is proposed for the streamwise turbulence in the outer region of a turbulent boundary layer.  The formulation pairs the logarithmic part of the profile (with a slope $A_1$ and additive constant $B_1$) to an outer linear part, and it accurately describes over 95\% of the boundary layer profile at high Reynolds numbers.  Once the slope $A_1$ is fixed,  $B_1$ is the only free parameter determining the fit. Most importantly, $B_1$ is shown to be proportional to the wake factor in the wall-wake law for the mean velocity, revealing a previously unsuspected connection between the turbulence and the mean flow.  
\end{abstract}

\section{Introduction}

The mean velocity distribution in the outer part of a turbulent boundary layer is often expressed as a combination of a logarithmic part and a wake component, as in
\begin{equation}
\frac{U}{u_\tau} = \frac{1}{\kappa} \ln{\frac{yu_\tau}{\nu}} +B + \frac{2\Pi}{\kappa} W \left( \frac{y}{\delta} \right),
\end{equation}
where $U$ is the mean velocity in the streamwise direction, $u_\tau = \sqrt{\tau_w/\rho}$, $\tau_w$ is the shear stress at the wall where $y=0$,  $\rho$ is the fluid density, $\kappa$ is von K\'arm\'an's constant, $B$ is the additive constant, and $\Pi$ is the Reynolds number dependent wake factor.  The wake function $W$ is taken to be a universal function of $y/\delta$, where $\delta$ is the outer layer length scale.   This wall-wake model was first formulated by \cite{Coles1956}, and it is an essential part of the widely-used composite profile derived by \cite{chauhan2007composite}.

\cite{Marusic_97} proposed a similar formulation for the streamwise turbulence intensity in zero pressure gradient turbulent boundary layers, given by
\begin{equation}
\overline{u^2}^+  =  B_1 - A_1 \ln \frac{y}{\delta_m} - V_g  \left( y^+, \frac{y}{\delta_m} \right) -W_g \left(\frac{y}{\delta_m} \right), 
\label{Perry1997} 
\end{equation}
where $\overline{u^2}^+ = \overline{u^2}/u_\tau^2$/. The model incorporates the log-law for $\overline{u^2}^+$ with constants $A_1$ and $B_1$ \citep{Hultmark2012, Marusic2013},  where $V_g$ is a mixed scale viscous deviation term, $W_g$ is the wake deviation term, and $\delta_m$ is a boundary layer thickness defined in a way that is similar to the Rotta-Clauser thickness.   It is typically 15 to 20\% larger than $\delta_{99}$, the 99\% thickness (further details are given in the Appendix).   
The wake deviation was given by 
\begin{equation}
W_g  =  B_1 \eta^2(3-2 \eta)- A_1 \eta^2 (1-\eta)(1-2\eta), 
\label{Wg} 
\end{equation}
where $\eta=y/\delta_m$.

\cite{pirozzoli2023outer} recently proposed an alternative model for the mean velocity distribution in the outer layer, given by a compound logarithmic-parabolic distribution 
of the type first suggested by  \cite{Hama1954a}.  That is,
\begin{eqnarray}
\frac{U_e-U}{u_\tau} & = & B - \frac{1}{k_0} \ln  \frac{y}{\delta_0}, \label{Hama1} \\
\frac{U_e-U}{u_\tau} & = & C \left( 1 -  \frac{y}{\delta_0} \right)^2, \label{Hama2} 
\end{eqnarray}
where $C$ is a constant,  and $U_e$ is the freestream velocity.   
Requiring the two velocity distributions to smoothly connect up to the first derivative 
yields the position of the matching point ($\eta_0=y_0/\delta_0$) and the additive constant $B$ in~\eqref{Hama1} 
as a function of $k_0$ and $C$,
\begin{equation}
\eta_0 =\frac{1}{2} \left( 1 - \left( 1 - \frac{2 }{C k_0} \right)^{1/2}\right), \qquad B = C (1-\eta_0)^2 + \frac{1}{k_0} \log \eta_0. \label{eq:patching}
\end{equation}
In what they called the classical case, \cite{pirozzoli2023outer}  found that with $\delta_0=1.6 \delta_{95}$ the best fit of the data was obtained
with $k_0 = \kappa \approx 0.38$, $C \approx 9.88$, so that $B=2.15$,
with the two distributions smoothly matched at $\eta_0=0.158$.  This compound logarithmic-parabolic distribution 
 fits the velocity distributions  well  down to $y/\delta_0 \approx 0.01$ (for Reynolds numbers based on displacement thickness greater than 2000). 

Here, we suggest a similar approach for the streamwise component of the turbulent stress.  That is, we propose a  compound representation given by
\begin{eqnarray}
\overline{u^2}^+ & = & B_1 - A_1 \ln \frac{y}{\delta_1} , \label{log} \\
\overline{u^2}^+ & = & b_1-a_1  \frac{y}{\delta_1} , \label{wake} 
\end{eqnarray}
where $a_1$, and $b_1$ are constants, and $\delta_1$ is the appropriate length scale for the outer layer.  In this formulation, there is no viscous deviation term. 
Requiring the two turbulence distributions to smoothly connect up to the first derivative 
yields the position of the matching point ($\eta_1=y_1/\delta_1$) and the additive constant $B_1$ in~\eqref{log} 
as a function of the other constants,
\begin{equation}
\eta_1 = \frac{A_1}{a_1}, \qquad B_1 = b_1-A_1 (1-\ln{\eta_1}). 
\label{patching}
\end{equation}
According to \eqref{wake},  $\overline{u^2}^+$ is zero when $y/\delta_1=b_1/a_1$, and so we impose one further constraint and set $b_1/a_1=1.05$ (which closely corresponds to the point where $y/\delta_{995} = 1$).  Finally, if we assume $A_1$ is a true constant ($=1.26$ for boundary layers according to \cite{Marusic2013}),  the only free parameter in our fit to the turbulence profile is $B_1$, which we will show to be Reynolds number dependent and proportional to the mean velocity wake factor  $\Pi$.

\section{Comparisons with data}

We now demonstrate the quality of the model by comparing it with  experimental and DNS data over a wide range of Reynolds numbers (see table~\ref{data}).    Before proceeding, we need to specify the particular length scale $\delta_1$ used to describe  the outer layer.   We have chosen $\delta_1=\delta_{99}$, for reasons made clear in the Appendix.   We also need to relate $Re_\theta=\theta U_e/\nu$,  where $\theta$ is the momentum thickness, to the friction Reynolds number $Re_\tau=\delta_{99}u_\tau/\nu$, in that not all data sets specify both.   This issue is also addressed in the Appendix.  

\begin{table}
\centering
{\begin{tabular}{lcccccc} \small
 & $Re_\theta$ & \ $Re_\tau$ \ & \ $b_1$ \ & \ $B_1$ \ & \ $\eta_1$ \ & \ Symbol \\
& & & & &  \\
\cite{DeGraaff2000} \hspace{1mm}   & 1430  & 541 & 3.56   & 1.05 &  0.372 & \blue{ \Large $\bullet$} \\
                                          & 2900  & 993 & 3.77   & 1.19  &  0.351 & \blue{ $\blacksquare$} \\
                                         & 5200  & 1692 & 4.33   & 1.57  &  0.306 & \blue{  $\blacktriangle$} \\
                                        & 13000  & 4336 & 4.83   & 1.94  &  0.274 & \blue{ $\blacklozenge$} \\
                                         & 31000  & 10023 & 4.67   & 1.82  &  0.276 & \blue{ \Large $\circ$} \\[2mm]
\cite{Fernholz_95a}          &  2573 & 866  & 4.10   & 1.42 &  0.323 & \ {\Large $\bullet$} \\
                                         &  5023 & 1692  & 4.34   & 1.59  &  0.305 & { $\blacksquare$}  \\
                                         &  7140 & 2375  &  4.80   & 1.92  &  0.276 & {  $\blacktriangle$} \\
                                         &  16080 & 5068  &  5.30   & 2.29  &  0.250 & { $\blacklozenge$} \\
                                         &  20920 & 6824  &  4.75   & 1.88  &  0.279 & \ {\Large $\circ$} \\
                                          &  41300 & 12633  &  4.75   & 1.88 &  0.279 & { $\square$} \\
                                          &  57720 & 18692  &  4.95   & 2.03 &  0.267 & { $\triangle$} \\
                                          &  60810 & 18362  & 4.95   & 2.03 &  0.267 & { \Large $\diamond$} \\[2mm]
 \cite{osaka1998re}           & 6040  & 1800 & 4.75  & 1.88 &  0.279 & \  \textcolor{Gray}{ \Large $\bullet$} \\[2mm]
\cite{Vallikivi2015_HRTF} & 8402  &  2622 & 4.90   & 1.99  &  0.270 & \  \ajs{ \Large $\bullet$} \\
                                         & 15121  &  4635 & 5.00   & 2.07  &  0.265 & \ \ajs{ $\blacksquare$} \\
                                         & 26884  &  8261 & 4.85   & 1.95  &  0.273 & \ \ajs{  $\blacktriangle$} \\
                                         & 46732  &  14717 & 4.80   & 1.92  &  0.276 & \ \ajs{ $\blacklozenge$} \\
                                         & 80579  &  25062 & 3.82   & 1.22  &  0.346 & \  \ \ajs{ \ \Large $\circ$ } \\
                                         & 133040  &  40053 & 3.57   & 1.06  &  0.371 & \   \ajs{ $\square$} \\
                                         & 234670  &  68392 & 3.23   & 0.845  &  0.410 & \ \ajs{ $\triangle$}  \\[2mm]
\cite{samie2018fully}        & 6252  &  1929 & 4.92   & 2.00  &  0.269 & \ \textcolor{OliveGreen}{ \Large $\bullet$} \\
                                         & 12913  &  3984 & 4.96   & 2.04  &  0.267 & \ \textcolor{OliveGreen}{ $\blacksquare$} \\
                                         & 26034  &  8032 & 5.10   & 2.14  &  0.259 & \ \textcolor{OliveGreen}{  $\blacktriangle$} \\
                                         & 47096  &  14530 & 4.88   & 1.98  &  0.271 & \ \textcolor{OliveGreen}{ $\blacklozenge$}\\[2mm]
\cite{Sillero2013}              & 6000  &  1848& 4.530   & 1.72  &  0.292 & \ \ {\bf - - - - } \\[2mm]
& & & & & \\[-3mm]
\end{tabular}}
\caption{Data sources and fitting parameters for $A_1=1.96$.  $B_1$ is the only free parameter, and $b_1$ and  the matching point $\eta_1$ are defined by \eqref{patching}.}
\label{data}
\end{table}

\begin{figure}
\begin{flushleft} {\small (a) \hspace {0.45\textwidth} (b)} \end{flushleft} 
   \centering
   \includegraphics[width=0.48\textwidth]{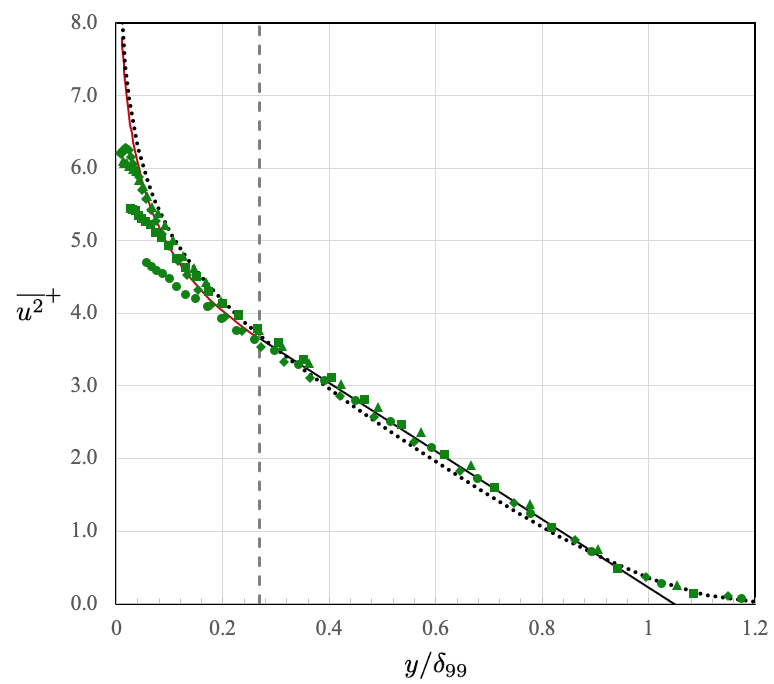} 
   \includegraphics[width=0.475\textwidth]{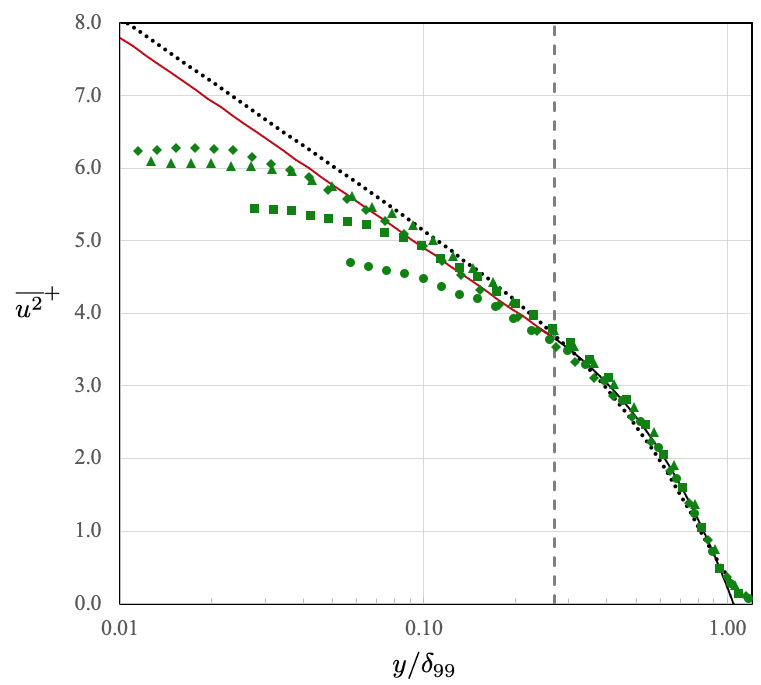} 
   \caption{Comparison to the experimental data of \cite{samie2018fully} for $Re_\theta=6252$--47096 ($y^+>100$, $A_1=1.26$, $B_1=2.00$).   (a) Linear scaling.   (b) Log scaling.  {\bf $\cdots \cdots$}, \eqref{Perry1997} (neglecting $V_g$);  \ajs{\bf ---------}, \eqref{log}; {\bf ---------}, \eqref{wake} (matched at $\eta_1=0.269$, vertical dashed line).   Symbols as in table~\ref{data}.  }
   \label{Samie}
 \end{figure}

To begin the analysis, we use the data by \cite{samie2018fully} ($6250 < Re_\theta < 47,$100).  Figure~\ref{Samie} demonstrates that, as expected from previous work, the log part with $A_1=1.26$ and $B_1=2.00$ is a good fit in the overlap region.  In addition, the linear part of the model describes the profile beyond the matching point very well over this range of Reynolds numbers,  except for the region $y/\delta_{99}> 1$ where a more gradual decline is observed.   At the highest Reynolds numbers, the compound formulation represents the data well for 95\% of the profile.   

For reference, we also plot \eqref{Perry1997} for the same values of $A_1$ and $B_1$ (using $\delta_{99}/\delta_m=0.81$).  We  neglected the viscous deviation term $V_g$, which leads to a positive offset of \eqref{Perry1997} with respect to  the compound in the logarithmic region.  Both fits work well beyond the logarithmic region, although it could be argued that the linear fit is a trifle more accurate for $0.2<y/\delta_{99}<0.9$.  In our analysis going forward, we will use  compound fit, primarily because the viscous deviation term in \eqref{Perry1997} appears to obscure some of the underlying trends as well as the comparisons between the turbulence and mean velocity profiles.

\begin{figure}
\begin{flushleft} {\small (a) \hspace {0.45\textwidth} (b)} \end{flushleft} 
   \centering
   \includegraphics[width=0.48\textwidth]{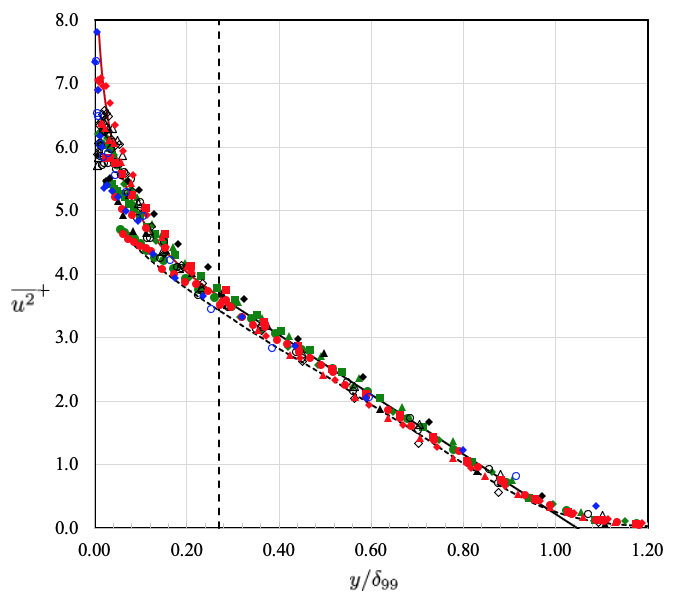} 
   \includegraphics[width=0.48\textwidth]{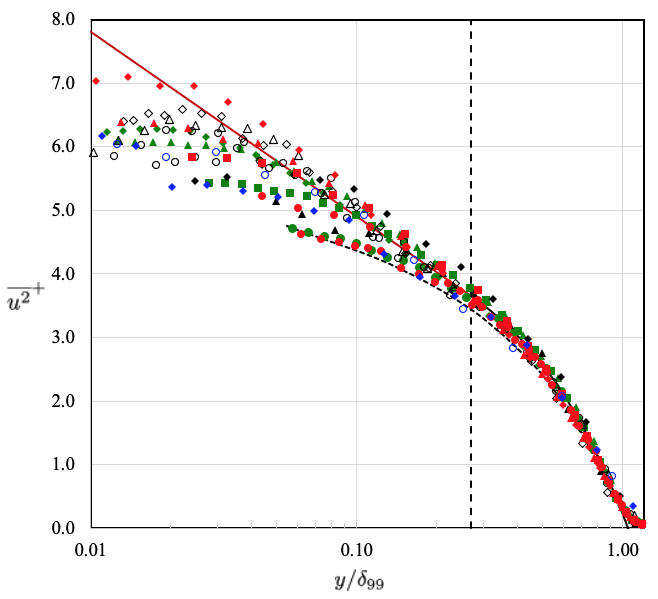} 
   \caption{Comparison to all the experimental data for $6000 \le Re_\theta < 60,000$ ($y^+>100$).  Symbols as in table~\ref{data}.   (a) Linear scaling.   (b) Log scaling.  {\bf ---------}, \eqref{wake};  \ajs{\bf ---------}, \eqref{log}.  $B_1=2.00$, distributions matched at $\eta_1=0.269$ (vertical dashed line). }
   \label{allhigh}
 \end{figure}

We now consider all the high Reynolds number data listed in table~\ref{data} over the range $6000 \le Re_\theta < 60,000$ (the Vallikivi {\it et al.\/} profiles for $Re_\theta>60,000$ will be dealt with separately).  The results are shown in figure~\ref{allhigh} for $B_1=2.00$.  Although there is some the scatter in the data, the compound fit works reasonably well using this value.  The agreement can be improved by using values of $B_1$ optimized for each profile, as listed in table~\ref{data}.

\begin{figure}
\begin{flushleft} {\small (a) \hspace {0.45\textwidth} (b)} \end{flushleft} 
   \centering
   \includegraphics[width=0.48\textwidth]{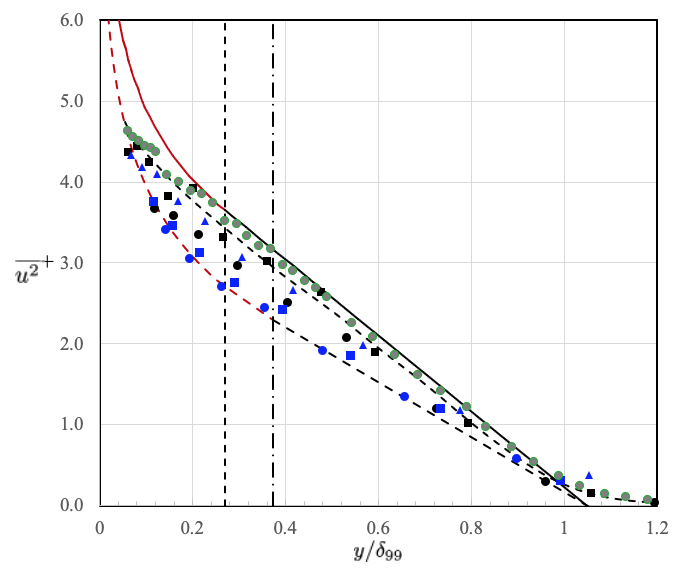} 
   \includegraphics[width=0.48\textwidth]{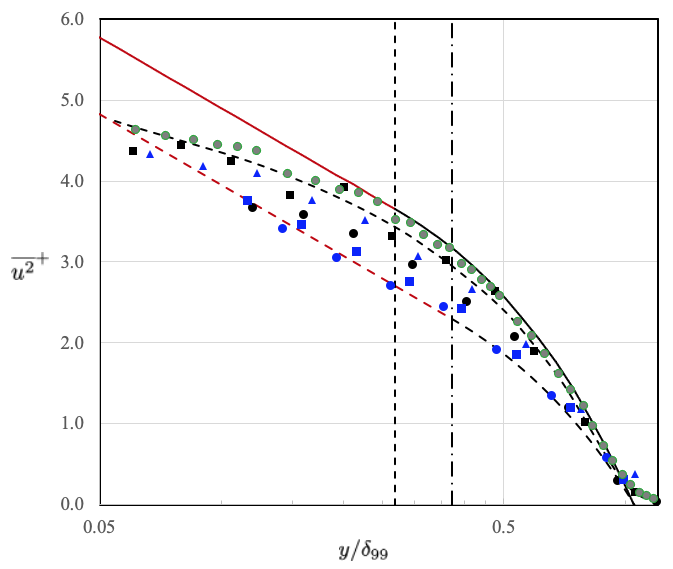} 
   \caption{Comparison to the experimental data for $Re_\theta \le 6040$ ($y^+>100$).  Symbols as in table~\ref{data}.   (a) Linear scaling.   (b) Log scaling.  {\bf ---------}, \eqref{wake};  \ajs{\bf ---------}, \eqref{log}.  $b_1=4.92$, distributions matched at $\eta_1=0.269$ (vertical dashed line).  {\bf - - - - -}, \eqref{wake};  \ajs{\bf - - - - - }, \eqref{log}.  $b_1=3.56$, distributions matched at $\eta_1=0.372$ (vertical dashed-dotted line). }
   \label{alllow}
 \end{figure}

The low Reynolds number data ($Re_\theta \le 6040$) are shown in figure~\ref{alllow}.  The compound fit is plotted for two cases, $B_1=2.00$ (the value used for the high Reynolds number data shown in figures~\ref{Samie} and \ref{allhigh}), and $B_1=1.05$ (chosen to match the lowest Reynolds number profile in the data set).  In order to match the in-between Reynolds number cases, $B_1$ was varied as given in table~\ref{data}.  Again, we see a very satisfactory fit to the data, even in this low Reynolds number range.

\begin{figure}
\begin{flushleft} {\small (a) \hspace {0.45\textwidth} (b)} \end{flushleft} 
   \centering
   \includegraphics[width=0.48\textwidth]{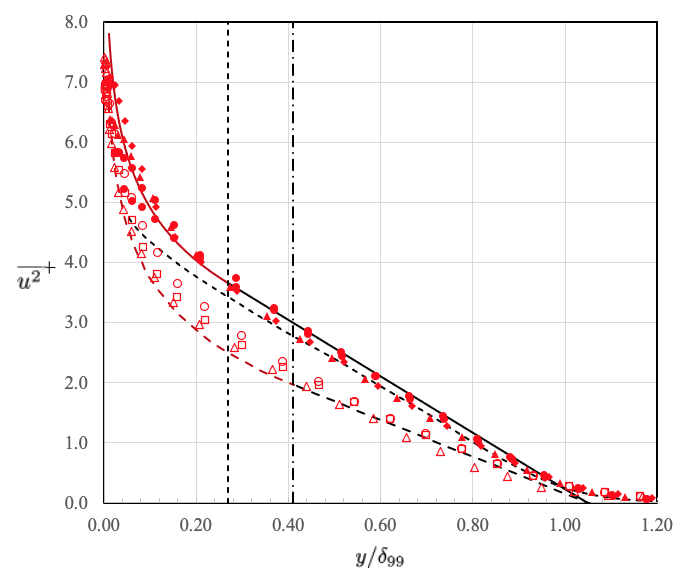} 
   \includegraphics[width=0.48\textwidth]{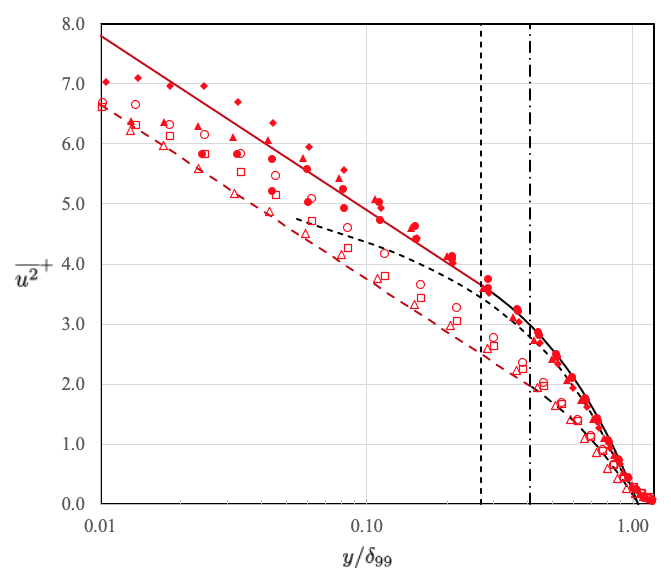} 
   \caption{Comparison to the experimental data for $Re_\theta > 60,000$ ($y^+>100$).  Symbols as in table~\ref{data}.   (a) Linear scaling.   (b) Log scaling.  {\bf ---------}, \eqref{wake};  \ajs{\bf ---------}, \eqref{log}.  $b_1=4.92$, distributions matched at $\eta_1=0.269$ (vertical dashed line).  {\bf - - - - -}, \eqref{wake};  \ajs{\bf - - - - - }, \eqref{log}.  $b_1=3.23$, distributions matched at $\eta_1=0.410$ (vertical dashed-dotted line). }
   \label{Valllkivi}
 \end{figure}

 We now consider the highest Reynolds number data, where $Re_\theta>60,000$.  There are only three profiles available,  at $Re_\theta=80,$579, $133,$040, and $234,$670  \citep{Vallikivi2015_HRTF}.  They are  unique, in the sense that no other turbulence data exist at comparable Reynolds numbers.  The compound fit is shown in figure~\ref{Valllkivi}.  Somewhat surprisingly, $B_1$ begins to decrease in a systematic way below its `high' Reynolds number value of 2.00 (see table~\ref{data}).  This result may indicate a problem with the data, but it may also reflect the  apparent connection between $B_1$ and $\Pi$.
 
\begin{figure}
   \centering
   \includegraphics[width=\textwidth]{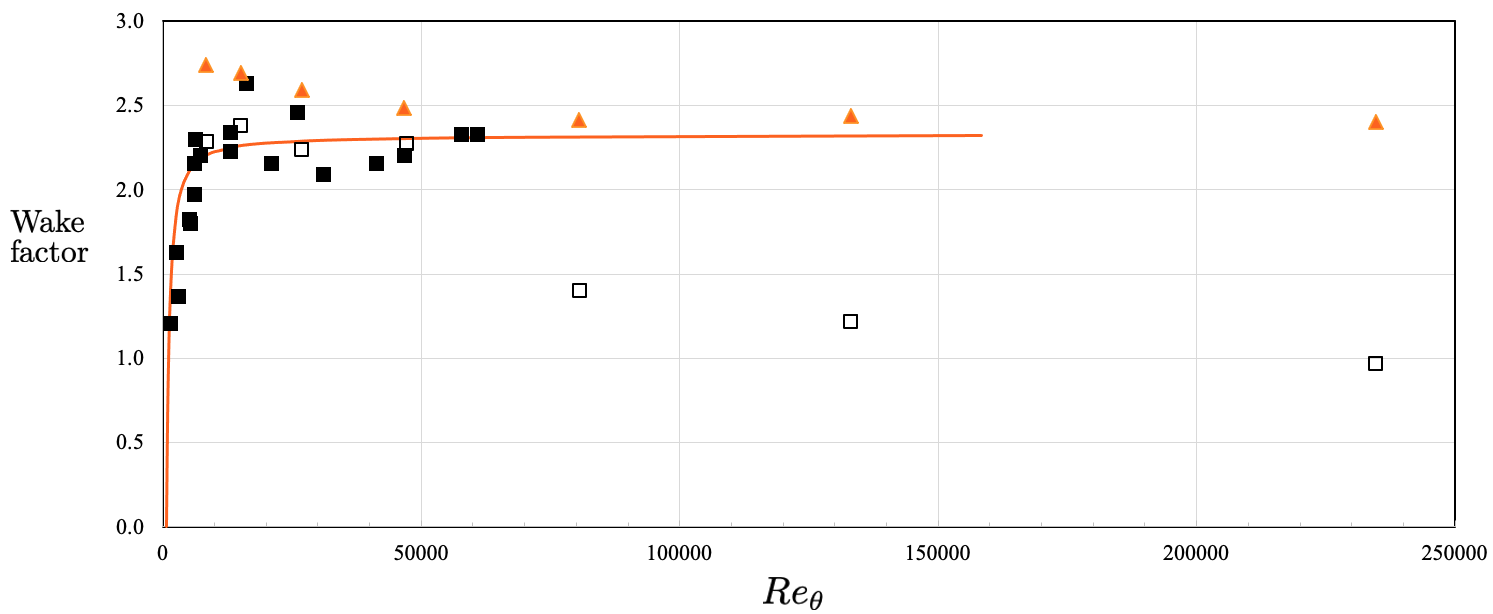}
   \caption{Wake factors versus $Re_\theta$.   {$ \blacksquare$}, 1.15$B_1$  (Vallikivi {\it et al.\/} 2015 values shown as {\small $\square$});  {\textcolor{RubineRed}{\bf --------}}, $2 \Pi/\kappa$ \citep{chauhan2007composite};  \textcolor{RubineRed}{$\blacktriangle$}, $2\Pi/\kappa$ \citep{Vallikivi2015_HRTF}.   }
   \label{piall}
 \end{figure}

The constant $B_1$ acts as a wake function for the turbulence profile, similar to the wake function $\Pi$ for the mean velocity profile.  In figure~\ref{piall}, we compare the Reynolds number dependence of $B_1$ with that of $2\Pi/\kappa$ (using $\kappa=0.384$; $B_1$ was scaled by an arbitrary factor of 1.15 to aid the comparison).  We see a clear similarity between the two wake functions for $Re_\theta <60,000$, and they appear to be proportional to each other by a constant factor.  

For $Re_\theta>60,000$, the  $B_1$ values decrease with Reynolds number, as noted above.  However, they still tend to follow the behavior of the corresponding $\Pi$ values.  Although the $\Pi$ values at lower Reynolds numbers for the Vallikivi data set  lie above the Chauhan correlation (which is similar to the \cite{Coles1956} distribution, not shown here), they consistently decrease with Reynolds number before becoming approximately constant, which is also the trend followed by $B_1$.  This observation is not  conclusive, but it helps to reinforce the notion that the two wake functions are linked.

\section{Conclusions and discussion}

The log-linear compound fit in $y/\delta_{99}$ proposed here for the streamwise turbulent stress $\overline{u^2}^+$ in the outer layer of a turbulent boundary layer works well over a wide range of Reynolds numbers.  For the log part of the fit we assumed that $A_1$, the slope of the log-law, is fixed at 1.26 (as given by \cite{Marusic2013}), and  the linear part of the fit was constrained to pass through zero at $y/\delta_{99}=1.05$.   As a consequence, the fit has only one free parameter, $B_1$, which acts like a wake factor.  

For low Reynolds numbers ($Re_\theta \le 6040$),  $B_1$  increases with increasing Reynolds number, attaining an approximately constant vaue of about 2 for higher Reynolds numbers ($6000 \le Re_\theta \le 60,000$).   At very high Reynolds numbers ($Re_\theta > 60,000$), $B_1$ begins to decrease with increasing Reynolds number, although this observation is based on very limited data.  

The behavior of $B_1$ closely follows the variation of the mean flow wake factor $\Pi$,  suggesting that the mean flow and the turbulence in the outer layer may be linked by a common mechanism related to the structure of turbulence.  This link is clear for the log-part of the formulation, in that the mean velocity and the turbulence distributions are connected through the attached eddy hypothesis \citep{PC82b, marusic2019attached}.  For the parabolic part of the mean velocity \eqref{Hama2} and the linear part of the turbulence \eqref{wake}, the link is still unknown.  One possibility is to consider the behavior of the ``detached'' (or Type B) eddies \citep{Perry1995}.  However, as they note, building this connection ``would be very complicated and would depend on the assumed shape of the representative eddies.''  Also, 
as \cite{hu_yang_zheng_2020} point out, ``Unlike the attached eddies, whose statistical behaviours are well described by the (attached eddy hypothesis), the detached eddies lack a good phenomenological model.''  In their statistically-based interpretation, the major contributions to the streamwise turbulence stress for $y/\delta >0.25$ are equally divided between the detached eddies and the Kolmogorov-scale eddies.  Understanding the physics that connects the mean velocity and the turbulence in the outer layer is clearly in need of further work.

\subsection*{Acknowledgements}
The authors would like to thank Sergio Pirozzoli and Jean-Paul Dussauge for their comments on an earlier draft.  

\vspace{3mm}
\noindent
{\bf Declaration of interests.}  The author reports no conflict of interest.

\section*{Appendix: Data analysis}

\begin{figure}
   \centering
   \includegraphics[width=0.48\textwidth]{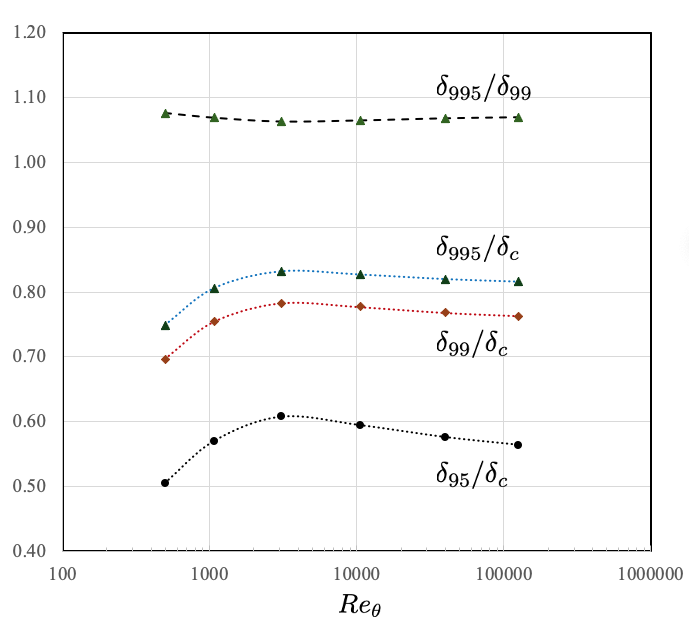} 
   \caption{Boundary layer thickness variations with $Re_\theta$, as found using the composite profile \citep{chauhan2007composite}.    }
   \label{thicknesses}
 \end{figure}

For the length scale used to describe  the outer layer, $\delta_1$,  there are a multitude of choices.  In examining the mean flow, \cite{pirozzoli2023outer} considered $\delta_0=1.6 \delta_{95}$, $0.28 \Delta$ (where $\Delta=(U_e/u_\tau)\delta^*$ is the Rotta-Clauser thickness), and $\delta_N=(H/(H-1))\delta^*$.   For the data in table~\ref{data}, 
\cite{Sillero2013}, \cite{DeGraaff2000} and \cite{Vallikivi2015_HRTF} used $\delta_{99}$, \cite{osaka1998re} used $\delta_{995}$, and \cite{samie2018fully} used $\delta_{c}$, where $\delta_c$ is the outer length scale adopted by \cite{chauhan2007composite} for their composite profile.   In order to compare data, we need a common standard, and  we will show that $\delta_1=\delta_{99}$ serves that purpose well.   To convert $\delta_{995}$ and $\delta_c$ to the matching value of $\delta_{99}$, we used the composite profile.  In figure~\ref{thicknesses}, we show how these various thicknesses compare.  

We used the results of \cite{Klebanoff1955_NACA} on the eddy viscosity.  His boundary layer thickness was about 1.15 times larger than $\delta_{99}$ \citep{Smits_Klebanoff}, and the data were scaled accordingly.

\begin{figure}
\begin{flushleft} {\small \hspace{0.06\textwidth} (a) \hspace {0.45\textwidth} (b)} \end{flushleft} 
   \centering
   \includegraphics[width=0.45\textwidth]{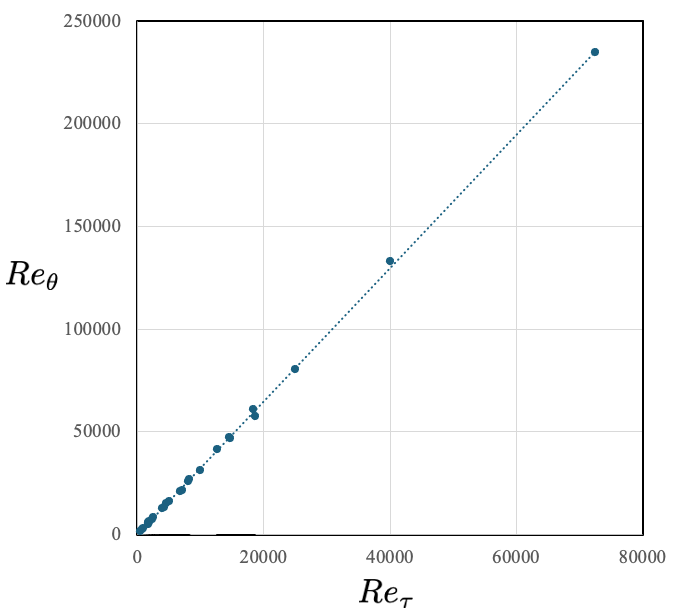} \hspace{2mm}
   \includegraphics[width=0.45\textwidth]{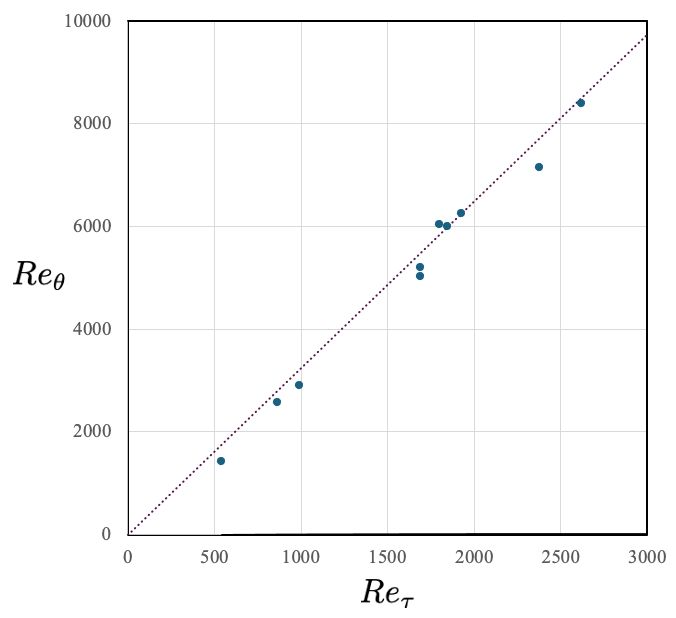} 
   \caption{Momentum thickness Reynolds number versus friction Reynolds number.   Dashed line,  \eqref{Reynoldsnumber}. (a) Full data set; (b) data for $Re_\theta < 10000$.   }
   \label{Reynoldsnum}
 \end{figure}

In addition, we need to relate $Re_\theta$ and $Re_\tau$, in that not all data sets specify both.  Here, we use  
\begin{equation}
Re_\theta = 3.241 Re_\tau,
\label{Reynoldsnumber}
\end{equation}
based on a fit to the available data ($R^2=0.9997$ for full data set).  See figure~\ref{Reynoldsnum}.

Finally, the $Re_\theta=234670$ profile by \cite{Vallikivi2015_HRTF} was corrected for an error in the 99\% thickness, which was smaller by a factor of 0.943 than the value originally reported.   This changed the profile, and the corresponding value of $Re_\tau$.  Also, the Fernholz profile at $Re_\theta=21410$ was not used since it has some obvious problems.


\end{document}